\documentclass[runningheads,a4paper]{llncs}
\pdfoutput=1

\usepackage{amssymb}

\usepackage{graphicx}
\graphicspath{{./graphics/}}

\usepackage[usenames,dvipsnames]{xcolor}
\definecolor{lightgray}{gray}{0.95}

\usepackage{tikz}
\usetikzlibrary{chains}
\usetikzlibrary{calc}
\usetikzlibrary{arrows}

\usepackage{listings}
\usepackage{url}
\urldef{\mailsa}\path|machens@tuhh.de|    
\newcommand{\keywords}[1]{\par\addvspace\baselineskip
\noindent\keywordname\enspace\ignorespaces#1}

\begin{document}

\mainmatter  

\title{Sandboxing for Software Transactional Memory with Deferred Updates}

\titlerunning{Sandboxing for Software Transactional Memory with Deferred Updates}

\author{Holger Machens}
\authorrunning{Sandboxing for Software Transactional Memory with Deferred Updates}

\institute{Institute of Telematics\\
Hamburg University of Technology\\
\mailsa\\
\url{http://www.ti5.tuhh.de}}

\toctitle{Sandboxing for Software Transactional Memory with Deferred Updates}
\tocauthor{Sandboxing for Software Transactional Memory with Deferred Updates}
\maketitle

\begin{abstract}
Software transactional memory implementations which allow transactions to
work on inconsistent states of shared data, risk to cause application visible
errors such as memory access violations or endless loops. Hence, many
implementations rely on repeated incremental validation of every read
of the transaction to always guarantee for a consistent view of shared data.
Because this eager validation technique generates significant processing costs
several proposals have been published to establish a sandbox for transactions,
which transparently prevents or suppresses those errors and thereby allows to reduce
the frequency of in-flight validations.

The most comprehensive sandboxing concept of transactions in software
transactional memory based on deferred updates and considering unmanaged
languages, integrates multiple techniques such as signal interposition, 
out-of-band validation and static and dynamic instrumentation. The latter
comprises the insertion of a validation barrier in front of every direct write
which addresses the execution stack of the thread and potentially results from
unvalidated reads.

This paper basically results from a review of this sandboxing approach, which
revealed some improvements for sandboxing on C/C++. Based on knowledge about the
runtime environment and the compiler an error model has been developed to
identify critical paths to application visible errors. This analysis lead to
a concept for stack protection with less frequent validation, an alternative
out-of-band validation technique and revealed additional risks from resource
exhaustion and so-called waivered regions inside transactions.

\keywords{software transactional memory, sandboxing}
\end{abstract}

\section{Introduction}

Transactional memory (TM) is the concept to apply a generic concurrency control
mechanism to critical sections in shared memory concurrent or parallel
applications. In applications using software transactional memory, the program
code contains transactional sections, which are properly instrumented to
redirect read/write access to shared data into the STM implementation. The concurrency
control of the STM then tries to establish a legal interleaving or overlapping
with concurrent accesses of other threads running transactions.

Many transactional memory implementations rely on optimistic concurrency
control with deferred updates, which basically means that transactions
speculatively work on possibly inconsistent data and every write attempt is
logged in a so-called write set until the end of the transaction (critical
section) is reached, its final state has been proven to be valid and the
write set is written back to its actual location of the shared data during
commit.

Generally, a transaction can be considered as an atomic transformation of
globally shared data from one consistent state into a new consistent state. In
this means a transaction is said to be valid if all data read (transaction's
read set) belongs to the same consistent state and does not contain updates of
unfinished commits from concurrent transactions. A violation of this rule
indicates an inconsistent state and requires the transaction to abort its
current run and restart from the beginning (transactional section entry). A
known method to validate the state of the transaction is to check each entry
stored in the read set for example by comparison to the current value present at
its original location in shared data.

A transaction working on inconsistent data is subject to several errors. The
most prominent of those errors are hardware exceptions such as memory access
violations (segmentation faults) and endless loops caused by inconsistent
termination conditions but there are more. To keep TM transparent those internal
errors have to be prevented or hidden from the application. A straight forward
guarantee is to entirely prohibit inconsistent states to be visible to other
transactions: a correctness criterion called
\textit{opacity}\cite{bib:CORRECTNESS}. Opacity can be achieved by incrementally
validating every read operation with the above validation method: a so-called
eager read validation, which unfortunately results in $\Omega(n^2)$ runtime
complexity, where $n$ represents the number of reads.

In contrast lazy validation reduces the overall validation effort of a
transaction but requires to deal with internal errors in terms of sandboxing to
hide them from the application. This paper reviews the currently most
comprehensive approach for sandboxing in unmanaged languages of Dalessandro and
Scott, generalises and specifies internal errors and their relationship, and presents
alternative improvements to sandboxing in regards to stack protection, resource
exhaustion and protection against endless loops or recursions.

The paper starts with an introduction to background knowledge on sandboxing
focusing on the approach of Dalessandro and Scott and considering an STM system
using deferred updates and lazy validation. The following section will present a
reviewed error model for doomed transactions, which is the basis for
improvements explained subsequently. Those improvements address the reduction of required
validation barriers through a different method to protect the data in the stack
and a general draft for a concept to reduce the risk of resource exhaustion.
Additionally, different alternatives of out-of-band validation for the
protection against endless loops will be specified. Prototypical implementations of the
sandboxing approach with different types of out-of-band validation have been
compared to eager validation using the Stanford
Transactional Applications for Multi-processing (STAMP benchmarks
\cite{bib:STAMP}).
The last section contains a discussion of the results gained through the
evaluation and the paper finishs with a conclusion.

\section{Background}
The only comprehensive sandboxing approach for C and C++ existing so far was
published by Dalessandro and Scott \cite{bib:SANDBOXING}. It is based on a
TM-aware C compiler \cite{bib:VELOX}, which automatically generates
instrumentation for those sections marked to be transactional. The programmer can define 
code blocks (see Listing \ref{lst:txcode}) to be transactional using some C++
language extensions such as specified in \cite{bib:CPPAPI}, which is a common
technique today and not related to sandboxing.

\begin{center}
\begin{minipage}{0.7\textwidth}
\begin{lstlisting}[caption={Example for a transactional
section},label=lst:txcode] 
/* local variables (thread-private) */
int a,b;
__transaction_atomic {
	
}
\end{lstlisting}
\end{minipage}
\end{center}

The basic STM system with lazy validation and deferred updates to be considered
here, stores every data read during the transaction in a so-called read set and
data to be written in a write set -- both can be considered as a log.
Considering transactions as atomic transformations of global data from one
consistent state into another, a validation step checks whether all read set
entries belong to exactly the same consistent state. It makes no significant difference
for sandboxing if the validation has to check the ownership of shared data
objects or just compares the value originally read with the current value at
original location. Due to the deferred updates, data to be written to globally
shared data by a transaction will get visible when the transaction has finished
with a commit only. Commits are mutually exclusive to other commits and
validations, which guarantees \textit{linearizability}
\cite{bib:LINEARIZABILITY}.
In contrast to globally shared data thread-private data such as local variables
are considered to be inaccessible to any other thread but the thread running
the transaction. Thus, thread-private data will be updated directly to achieve better
performance. However, in case of a rollback the transaction has to revert direct
updates of private data.

A transaction using lazy validation runs into inconsistencies when reading
shared data belonging to different consistent states. This can be best explained
using the privatisation scheme as example use case. Privatisation basically
means to remove shared data from access of other threads to allow private access
without concurrency control by the privatising thread. Consider for
example a shared pointer to some shared data object.
To privatise the data object the privatising thread can set the pointer
to zero (\texttt{NULL}), indicating that the object is no longer available to
the public.
Now consider a case, where a transaction $T_r$ reads the pointer, another
transaction $T_p$ privatises the pointer concurrently and deletes the
formerly addressed data object and finally the transaction $T_r$ tries
to read the no longer existing data object. The result may be a memory access
violation producing a segmentation fault signal or just inconsistent data,
which may result in subsequent errors such as an endless loop due to inconsistent
data in the termination condition. Hence, a transaction working on inconsistent
data is doomed to suffer from internal errors. To keep transactional memory
transparent, these internal errors are not allowed to affect the ordinary
application behaviour.

The sandboxing concept is generally, to either catch and undo errors of doomed
transactions or prevent them by additional validation and possible rollbacks. In
this regard, Dalessandro and Scott considered the following list of
non-transparent events in transactions to be covered by their sandboxing
approach:

\begin{itemize}
  \item \textit{In-place Stores}: In-place stores are all writes performed by a transaction which are not
  redirected into the write set, i.e. direct updates in contrast
  to deferred updates. Direct writes based on inconsistent reads can result in
  all kinds of severe errors. Just consider an inconsistent address to be used
  in a direct write attempt: It can address private or shared data or even
  memory areas never allocated and cause all kinds of subsequent errors.
  \item \textit{Indirect Branches}: Much like indirect references,
  indirect branches refer to a target address, which is dynamically computed
  (unknown at compile time).
  Indirect branches are used in virtual function calls, large switch statements
  and \texttt{goto}s with computed target address.
  Obviously, branching to an inconsistent target address will severely alter the
  program behaviour and may even cause the transaction to leave the
  transactional section temporary or finally without commit. Branching to
  an instruction without following the intended control flow of the program
  already violates against transparency. Furthermore, executing
  non-transactional code during a transaction can cause all kinds of errors inside and outside of the STM system.
  \item \textit{Faults}: Interrupts triggered by the processor on detected
  errors are called faults or exceptions. Those faults are for
  example memory access violations known as segmentation faults, floating point
  exceptions such as division by zero and others. The effect of many
  operations on inconsistent data can be these faults that become visible to
  the application in most cases as an immediate program termination.
  \item \textit{Infinite Loops and Recursion}: Loops and recursive function
  calls always require some termination condition. Inconsistency of the data
  used in the logical expression can cause the condition to be false forever and
  the transaction gets stuck inside the recursion or loop.
\end{itemize}

They integrated a special extension in the TM-aware compiler to insert
pre-validation hooks in front of \textit{dangerous operations}. Dangerous
operations are basically in-place stores, whether on stack or shared data, and
indirect branches which result from shared reads according to the data flow.
Every read from shared data is considered to be inconsistent unless it was
validated. If the data read is used in a subsequent dangerous operation, it
requires a pre-validation to prevent possible errors. Hence, the compiler
extension analyses the data flow and instruments every in-place store and
indirect branch with pre-validation code which are the first accessing data read
from shared location in terms of data flow. The inserted code decides at runtime
whether a validation of the reads in a particular data flow branch is necessary
or not and aborts in case of inconsistencies.

Faults are validated in special POSIX signal handlers. If the transaction is
proven to be valid, the fault is raised to the application, otherwise it is
suppressed and results in a rollback.

A timer-based out-of-band validation is applied to escape from infinite loops and
recursions. It is simply triggered by a POSIX timer, which raises an interrupt
with a frequency between $1$Hz and $100$Hz dynamically adapting to the frequency
of detected errors. The interrupt is processed in another signal handler which
performs the validation.

For each indirect function call (i.e. via a function pointer) the compiler
inserts a lookup of the corresponding transactional function. If only a
non-transactional function is found, the STM switches to its \textit{irrevocable} mode, which implies a
pre-validation and mutually exclusiveness to all other running transactions
until the transaction finishs. Hence, an invalid function pointer without a
match to a transactional function is pre-validated implicitly. This also
prevents non-transactional functions, which includes almost all system and
library functions to be called with inconsistent parameters.

They also considered intentionally inserted non-transactional code sections such
as so-called \textit{waivered code} sections or \textit{pure functions} (cf. GCC
4.7) by additional pre-validation.

These four mechanism pre-validation of in-place stores and indirect branches,
catching faults, out-of-band validation and function pointer validation properly
sandbox all non-transparent events mentioned above. The main difference to eager
validation is that the validation of the read set is moved to the location of
potentially or actually occurring errors or the end of the transaction. That
means in turn: the more potentially dangerous operations exist inside the
transactional section the more pre-validation hooks will be executed and the
overall validation effort will be similar to eager validation in the
end. 

In-place stores are the most frequent events considered to be potentially
dangerous. In contrast, indirect branches from in-place stores are quite rare and
we consider the remaining errors possibly resulting from in-place stores to be
handled by the other mechanisms anyways.

The possible reduction was our motivation to review the approach of Dalessandro
and Scott by an analysis of the cause and effect relationships of the errors in
doomed transactions under consideration of a more realistic model of computing
in concurrent application using transaction on modern hardware.

\section{Model of Computing}

Dalessandro and Scott considered a more generalised model of computing to be
supported by their sandboxing method: A thread reads and writes data shared or private and
follows some control flow, which contains branches. Reads and writes of a thread
are in relationship in means of a data flow. Based on this model, their approach
is possibly the only viable answer. But actually a C/C++ application compiled
for a modern computer system and using STM with deferred updates defines a
certain model of computing, which reduces the potential risk of in-place stores.

\begin{figure}[ht]
\begin{center}
\includegraphics[width=.6\textwidth]{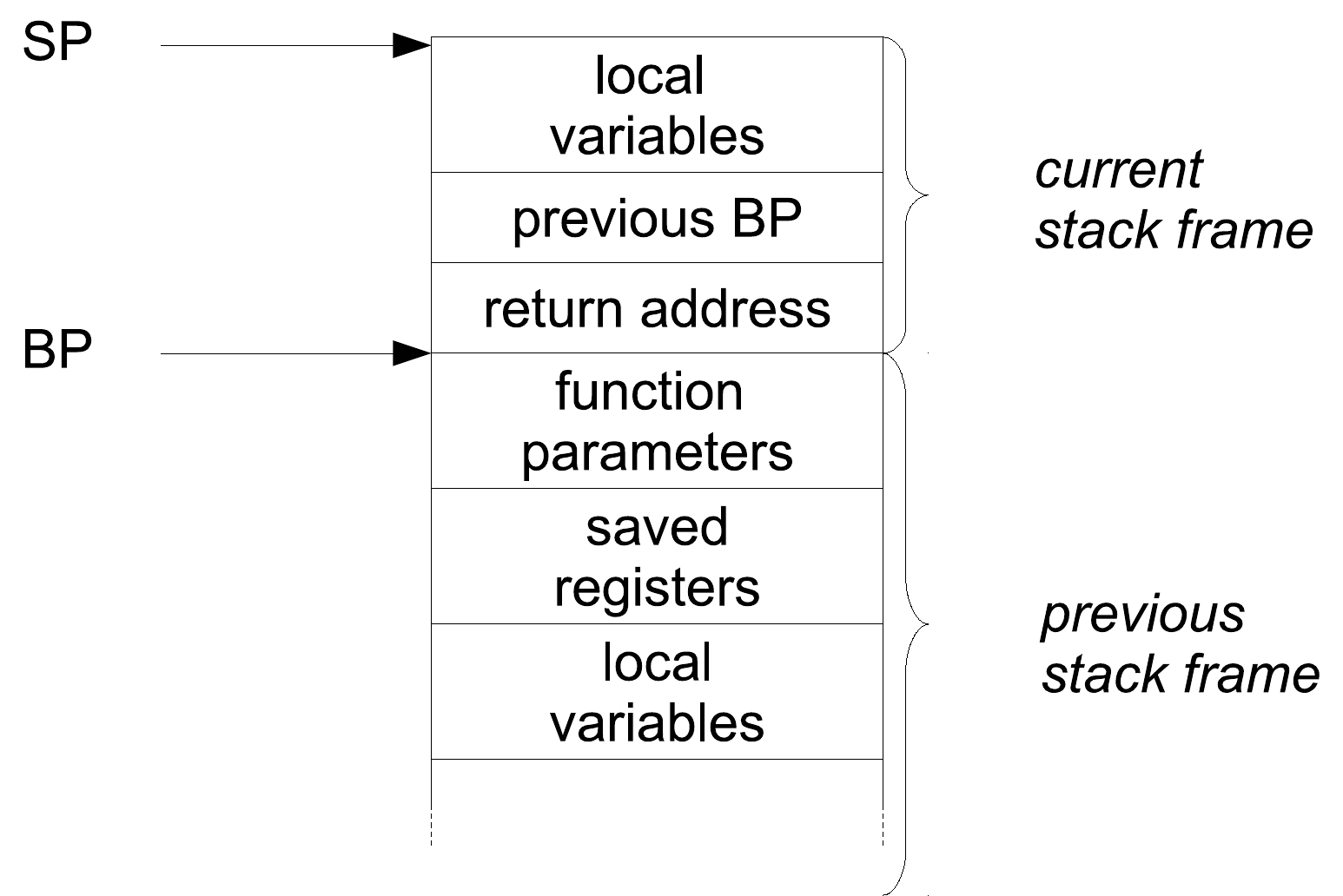}
\caption{Execution stack of a thread}
\label{fig:stack}
\end{center}
\end{figure}

A concurrent program runs multiple application threads, which execute
instructions. We consider a transaction to be executed by a single thread only.
Thus, a thread will never spawn another thread whilst executing a transaction.
Each thread uses an execution stack to store private data and occupies a
processor (or core) when running. The core uses registers to operate, holding
data for certain purposes:

\begin{itemize}
  \item \textit{Instruction pointer} or \textit{program
  counter}:   A reference on the next instruction in the application code to be
  executed by the thread, incremented with each instruction and updated
  whenever the control flow branches to another location in the code (jump and
  call or return from function).
  \item \textit{Stack} (SP) and \textit{base pointer} (BP):
  References to the top and the bottom of the stack frame on the execution stack
  currently used by the thread to execute a particular function. These registers
  are used to access the data stored inside the stack frame. 
  \item \textit{General purpose registers}: Those registers are available for
  arbitrary purposes:
  \begin{itemize}
  \item Instruction parameters: Machine code instructions expect parameters to
  be passed through particular registers as specified.
  \item Temporary variables: Most registers have no specific purpose and are
  free to be used by the thread to temporary store data e.g. for
  local variables or parameters of functions.
  \end{itemize}
\end{itemize}

The execution stack (see Figure \ref{fig:stack}) is especially needed to store
the local data of recursive function calls in particular and has become a
standard mechanism to implement function calls on most platforms. An execution
stack stacks so-called \textit{stack frames} for each function not yet finished
by the thread. The stack frame contains data such as the return address, a
reference on the previous stack frame (BP), the function parameters and local
variables. The location of this data in a certain stack frame is fixed once the
application was compiled. All C/C++ compilers for the majority of general
purpose computer architectures generate code, which complies with this general
model of computing in regards to the stack, function calls and registers
mentioned here.

When considering deferred updates, the STM system limits in-place stores in
active transactions (not in a waivered section) to data on the stack which is
guaranteed to contain thread-private data only. Any other variable on the stack
which is not guaranteed to be private may be shared with another thread and is
by definition not allowed to be written directly in a deferred update STM.
For example a variable that has been defined in the \texttt{main}
function\footnote{entry function of every C program.} of an application is
stored on the stack too and can of course be shared with other threads started
subsequently. In contrast, a local variable of a function executed inside a
transactional section cannot be shared with other threads, because it will no
longer be available when the function returns. The STM is responsible to decide
whether a write addresses a private or potentially shared location no matter if
it uses lazy or eager validation. To achieve this, the STM determines which part
of the stack is currently \textit{transaction local} and applies an appropriate
filter to all write attempts. Thus, the remaining risk of in-place stores is
limited to modifications of sensitive data inside the transaction local part
of the stack.

\section{Error Model}

For our error model we tried to identify the actual application visible
anomalies which are finally irrecoverable. The following list covers all such
application visible anomalies caused by doomed transactions in C/C++
applications considering the model of computation given in the previous section.
The list renders that of Dalessandro and Scott for the most part but we added
another class of anomalies (resource exhaustion) and chose another sectioning to
identify cause and effect relationships later by backtracking.

\begin{enumerate}
  \item \textbf{Hardware exceptions}: Those are the faults/interrupts generated
  by the hardware on these exceptions:
  \begin{enumerate}
    \item Memory access violations while using
    inconsistent pointers or pointers to inconsistent data
    such as deleted data. 
    \item Arithmetic exceptions such as division by zero or
    overflows.
    \item Illegal instruction errors when entering a memory section that does 
    not contain executable code for example caused by mistaken jumps or
    inconsistent modifications to code, considering self-modifying code.
  \end{enumerate}
  \item \textbf{Incomplete transaction exit}: A transaction may accidentally
  escape the critical section without a proper commit. Not committed writes are no longer
  visible to the thread, the thread can be asynchronously rolled back leaving shared
  data in an inconsistent state due to the lack of transaction support and the
  next entry of a transaction is interpreted as a nested transaction resulting 
  in more errors. Reasons for these escapes are inconsistent function pointers 
  or jump targets such as the return address saved on stack or dynamically
  computed jump targets of \texttt{switch}-\texttt{case} blocks or
  \texttt{goto} statements.
  \item \textbf{Computation of non-transactional code}:
  A transaction may accidentally enter non-transactional code such as an
  uninstrumented function, while inside
  the critical section. Again, uncommitted data is not visible and unexpected 
  rollbacks may occur leaving shared data in an inconsistent
  state as in case 2 above. Reasons are the same as for incomplete transaction
  exits but here the thread will at least return to the transactional code.
  \item \textbf{Infinite loop/recursion}: Inconsistent termination conditions
  may result in endless loops or recursions. This endless loop or recursion may as
  well contain no instrumentation so the STM cannot get aware of this state.
  \item \textbf{Resource exhaustion}: Inconsistent parameters to allocation
 requests for resources such as main memory or space on a hard drive (files) may
 result in an exhaustion of that resource affecting even unrelated applications
 on the system.
\end{enumerate}

\begin{figure}[ht]
\begin{center}
\includegraphics[width=1.0\textwidth]{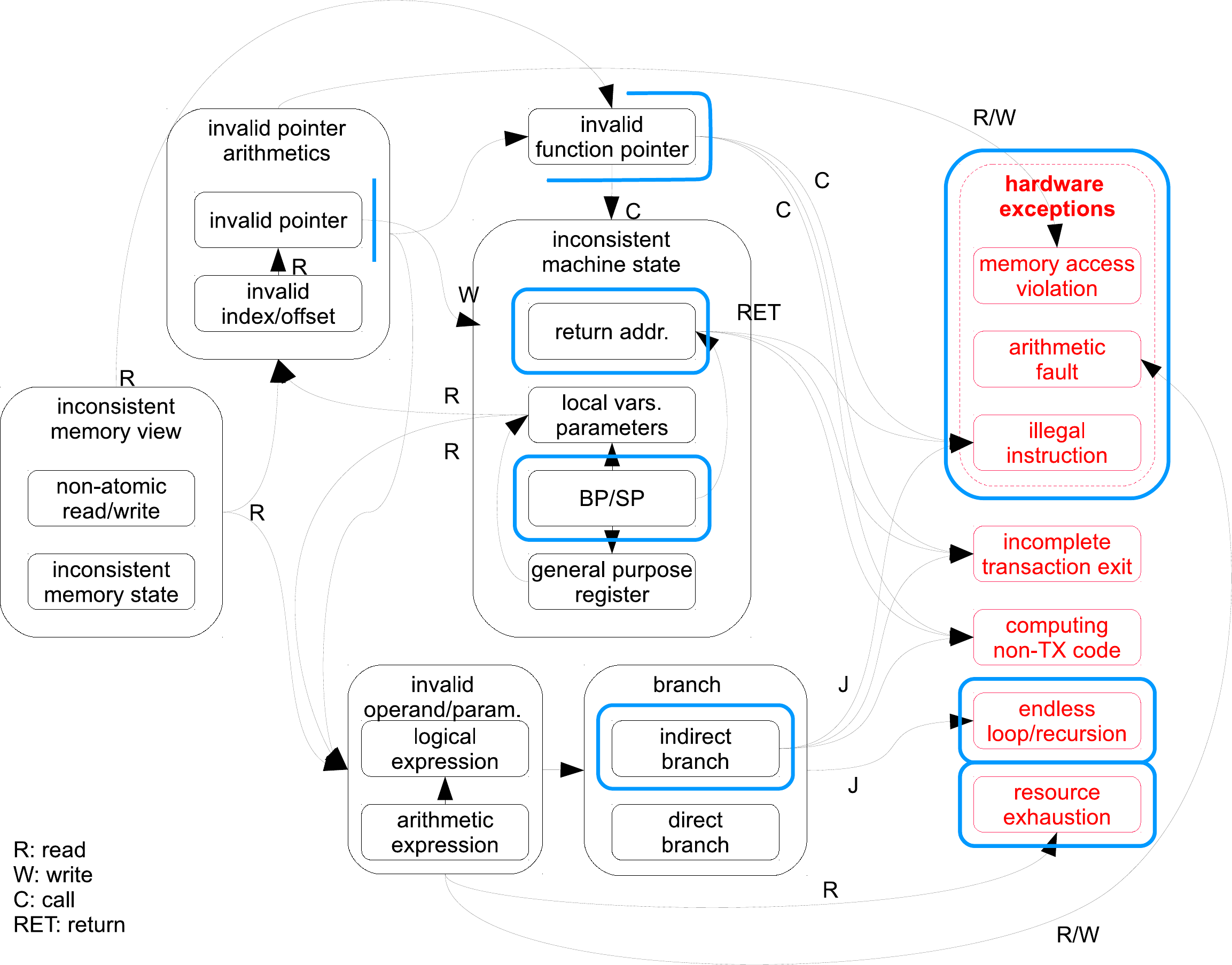}
\caption{Error model for doomed transactions}
\label{fig:error}
\end{center}
\end{figure}

The error model given in Figure \ref{fig:error} provides an overview of
the dependencies between effects of doomed transactions to the left and
application visible anomalies marked in red to the right (the blue
lines will be explained later).
Considering inconsistent memory states and non-atomic reads/writes as root
causes for all possible transparency violations the arrows point towards
subsequent errors. Each dependency is associated with one or more low level instructions
such as read (\texttt{R}), write (\texttt{W}), jump (\texttt{J}) and function
call (\texttt{C}) or return (\texttt{RET}). To reduce the complexity of the
graph we have removed all dependencies which just result in cycles such as
invalid operands to an arithmetic expression causing more inconsistent
data. The following paragraphs will explain the characteristics of the
effects of inconsistencies depicted in Figure \ref{fig:error}.

\begin{description}
\item[Inconsistent Memory View] This box represents the view of the transaction
on the shared data in memory, which may differ during the course of execution.
We generally differentiate between an inconsistent memory state and inconsistent
data due to non-atomic reads/writes. 

\begin{description} 
\item[Non-atomic Reads and Writes] Read of native data types is not guaranteed
to be atomic. For example a read of a 64 bit integer which spans over different
cache lines may overlap a write of the same integer and results in an
inconsistent read. Direct effects of inconsistent reads are invalid pointers,
invalid indices in arrays or invalid operands in general.

\item[Inconsistent Read Set] An inconsistent read set exists when a
transaction reads data from different serialisation orders. For example 
a transaction reads two variables that are updated by another transaction. 
If the transaction reads the first variable before it gets updated and the 
second from the updated state, both reads do not
belong to the same consistent memory state. Instead the reads reflect an
inconsistent memory state where just one variable has been updated.

\end{description}

\item[Invalid Pointer Arithmetic] This category addresses all kinds of errors
resulting from invalid pointer arithmetic such as use of pointers on freed
memory, invalid pointers in general and invalid indexes or offsets used in
pointer arithmetic. Memory referenced by an invalid pointer may be non-existent
or freed by another thread. This can result in a memory access violation
detected and signalled by the hardware through an interrupt. The same memory
area might have been issued to another thread in the meantime or the pointer just
refers to some completely different location and the doomed transaction just
accesses chaotic data.

Invalid pointers are the result of inconsistent reads on shared data only.
Read of private data, such as local variables on stack, can be considered to be
valid as long as reads from shared data are consistent. Direct write access via
invalid pointers obviously results in unpredictable effects on shared and local
data affecting the whole application.

In the programming language C, indexes in arrays and offsets to pointers are
actually the same and treated as input to indirect references (i.e. pointer
arithmetic). Once computed, both can have the same effects as invalid pointers
in the first place. Because, STM implementations usually grant direct access to
data in stack such as local variables, inconsistent indexes or offset may
result in uncontrolled modifications in the stack too, which can cause an
inconsistent machine state (see below).

\item[Invalid Operands] Invalid operands are the result of inconsistent reads
from shared data, data referenced by invalid pointers or inconsistent machine
states (see below). Invalid operands may affect logical or arithmetic
expressions or computed jump targets causing access to non-transactional code,
arithmetic faults, resource exhaustion or infinite loops or recursions.

\item[Inconsistent Machine State] An inconsistent machine state may be caused by
an invalid function pointer or inconsistent stack content for example caused by
off-by-one errors using invalid indexes or offsets on local variables or arrays.
As introduced with our model of computing, the base
pointer (BP) of the previous stack frame may be stored on the stack. For example an
inconsistent write access to an array on the stack can cause the BP to be
overwritten. If the thread restores the previous stack frame when leaving the
function, it uses this erroneous BP and the whole stack frame is wrong.
Just as BP the return address can be modified, too. When leaving the
function the thread consequently jumps to this address, which may point to any
position in the virtual address space resulting in all kinds of errors such as
entering non-transactional code or even sections that contain no proper
instructions at all. Before entering a function the thread also stores several
registers on the stack to be restored on return. Those may be affected by
inconsistencies the same way as BP and the return address above, resulting
in inappropriate inconsistent local variables when restored.

\end{description}

\section{Sandboxing Concept}
\label{sec:CONCEPT}

A remaining disadvantage of the approach of Dalessandro and Scott is the
pre-validation of in-place stores. 

Because in-place stores are restricted by the STM to private data on the part of
the stack belonging to the transaction as defined by our model of computing,
this is the only memory region to be protected against inconsistent writes.
Inconsistent writes are dangerous only if the resulting effects cannot be
suppressed. Inconsistency of local variables or parameters is harmless as long
as the effects can be handled: Hardware exceptions can be handled by validation
in signal handlers, endless loops can be terminated by an out-of-band
validation and resource exhaustion can be handled too as we will explain later.
In contrast, modifications of BP or the return address can finally result in
an incomplete transaction exit or computing of non-transactional code, when a
the thread returns from a function. Validation of the return address, before
returning from the function is one approach to deal with it, but we
went for another.

The location of local variables and parameters in the stack frames is fixed once
the application was compiled and referenced indirectly via SP or BP at runtime.
Thus, as long as we can guarantee BP and SP to be valid, directly addressed
in-place stores will always address the right location and cannot cause
inconsistent returns. But indirect access via pointer arithmetic or arrays
(which is the same in C) might still harm BP and the return address inside
the stack and will require pre-validation when targeting stack contents.
Thus, the only in-place stores to be validated are those via pointers into
backups of the machine state inside the stack, particularly the BP and the
return address field. This dependency is depicted in Figure \ref{fig:error} by
the write access relationship between \textit{invalid pointer arithmetic} and
\textit{inconsistent machine state}.

As we know, access to locations outside the context of the transaction on the
stack has to be filtered by the STM anyway. A TM aware compiler does not
instrument access to local variables or function parameters though, because they
are considered to be private. Indirect access via non-constant pointers is
unpredictable at compile time because the referenced address is calculated at
runtime and may address private or shared data. Therefore this kind of indirect
access always requires instrumentation and the STM has to evaluate whether the
access has to be handled transactional or non-transactional. Thus, we can use
the same instrumentation to identify access to the critical parts in the stack
such as BP and the return address. Because access to this critical content
cannot be part of a C program, unless it uses embedded assembler blocks, which
are not allowed inside transactional sections (see \cite{bib:CPPAPI}), we can
consider them as non-intended by the programmer. Because access to critical
parts in the stack is invalid per definition, it does not need a full
validation of the read set and immediately aborts the transaction, which is the
main difference to validation of the return address proposed earlier. In case of
an actual application error this can result in an endless loop. For debugging
purposes the transaction could check if it actually works on consistent data
(i.e. if no other thread performed a commit) and terminate the process to
interrupt the endless loop. But actually such an application error can result
in an endless loop anyway and we don't have an advantage adding the validation
here.

The interception points in the development of application visible errors in
the graph in Figure \ref{fig:error} are depicted by blue lines. The machine
state protection by validation of invalid pointer arithmetic related to the
stack represents the main difference to the sandboxing approach of Dalessandro
and Scott and suppresses a lot of subsequent errors.

We do not see more effective alternatives to concepts such as out-of-band validation
to terminate endless loops/recursions and validation of function
pointers or signal interposition to handle hardware exceptions, and will use
them in our approach as well. However, function pointers and signal
interposition does not need a full validation in our approach. 

There are just two reasons for a dangerous signal such as segmentation faults or
floating point exceptions to occur:
\begin{itemize}
  \item The transaction is working on inconsistent data and requires a
  rollback to hide the signal.
  \item The application is actually erroneous in which case the signal will
  cause a termination of the process.
\end{itemize}
In a correct application the second case will never occur and thus we decided
to optimise the first case. If the globally shared timer indicates that another
transaction has committed and potentially modified globally shared data, the
signal is considered as invalid and the transaction aborts immediately without
a full validation of the read set. In case of an actual application error the
signal will be produced again while the transaction is valid and finally
terminate the process.

The unsuccessful lookup of transactional clones for function pointers can be
treated similar as occurring signals and access to sensitive parts of the stack.
Those are by definition invalid and must be errors of the application or the
result of an invalid state of the transaction. Thus, on unsuccessful lookups the
transaction aborts considering the state to be invalid. Again, for debugging
purposes a validation helps here but is not needed for correctly implemented
applications.

For out-of-band validation two different approaches exist. Dalessandro and Scott
used a timer-triggered validation, which causes the thread executing the
transaction to interrupt its work and validate. Thus, the effort of out-of-band
validation considered out-of-band is actually in-band of the transaction.
And because the timer triggers at certain times only there will always be a gap
between occurrence and detection of an inconsistency. An alternative proposed by
Kester et al. \cite{bib:DELEGATEVALIDATION} earlier is to delegate the out-of-band
validation from the \textit{leading} thread executing the transaction to a helper thread.
The helper thread constantly checks the read set of the leading thread and
globally shared data and delivers signals to the leading thread on detected
inconsistencies. This approach removes the effort from the \textit{leading}
thread and detects inconsistencies instantly. But the additional concurrent access to shared data
causes the application to significantly slow down, as we will demonstrate in the
evaluation later. That was probably the reason why Kestor et al. used a
multi-processing architecture with hyper-threading, which shares the first level
cache beyond the two \textit{hyper}-threads and thereby reduces the cache
contention between leader and helper thread.

For a more hardware independent approach, we developed a third kind of
out-of-band validation, which reduces the communication effort between leader
and helper thread. Instead of validating the read set of the leading thread the
helper thread virtually executes an exact clone of the transaction on its own
and applies eager validation instead. The helper thread is not meant to commit
the transaction at any time and will not write to any location but its own
stack. Because the helper thread starts its execution of the cloned transaction
on the same state of globally shared data it will consequently take the same
control paths, and develop the same read set over time as the leader. As soon as
the helper thread detects an inconsistency due to a validation in a read
attempt, it notifies the leader about it and restarts on the new state of
globally shared data. On notification the leader will abort as well unless it
has already reached the end of the transaction, did its own lazy validation and
aborted or committed, consequently. If the helper successfully reaches the end
of the transaction it checks whether the leader has already finished or not. If
not it keeps validating its own read set until the leader has finished or a
conflict occurs, which is handled the same way as above. This approach combines
the advantages of lazy and eager validation, but it also introduces a
disadvantage in respect to waivered sections due to additional parallelism
inside the transaction to be explained later.

Resource exhaustion can be either prevented through pre-validation or resolved
on demand. Because pre-validation would decrease the advantage of sandboxing
again, we have analysed alternatives to solve the exhaustion on demand.
Most of the possible resource exhaustion incidents will be pre-validated anyway,
because the system library functions are usually not declared to be
transactional and the STM has to switch to the irrevocable mode, which implies a
validation. But there are functions needed to support transactional behaviour
and acquire system resources such as \texttt{malloc}\footnote{Please note, that
we consider malloc to be supported by the STM system here. Thus, the transaction
will not turn into irrevocable mode to execute it.} for memory management and we
can think of other resources such as file system space to be handled
transactional as well.

We will use \texttt{malloc} as an example to explain the issues in prevention of
resource exhaustion. Actually, it is easy to detect the lack of a memory: The
\texttt{malloc} function returns an error code, which indicates that no more
memory is available. System memory is limited by the available main memory and
possibly existing file system space to swap out pages. Thus, the system memory
is shared by all processes and all threads and the incident may be shared by
multiple transactions and even other applications in the system, consequently.
Of course the invalid transaction will detect this inconsistency later and resolve
it by a rollback releasing the allocated memory. But it causes a lot of
interference in the system even if there is still memory available: The system
starts to swap pages of other threads and the invalid transaction causes the
whole system to slow down. Thus, there is a fairness problem that can be solved
by a system wide control mechanism only that has to be integrated with the
subsystem which controls the resource such as the memory management in our case.

Although it is technically possible, the development of a system wide control
mechanism to protect against effects of resource exhaustion is beyond the topic
of this paper. Hence, we decided to use a hybrid method for memory allocation, which
applies pre-validation in cases where a given limit of requested memory by the
thread is exceeded, only. This still does not solve the actual problem but it
reduces the risk of resource exhaustion sufficiently to allow a proper evaluation 
of the sandboxing approach considering a properly dimensioned main memory.

The other remaining issue is caused by computed jump targets, which might result
in computation of non-transactional code. We have found no alternative
solution to pre-validation in this case either. However, the use of such
dynamically computed targets in indirect branches by the programmer (i.e.
\texttt{goto}) is very rare, and indirect branching in large switch
statements is actually not necessary. Thus, we decided to prohibit them
entirely in transactional sections. However, virtual function calls have to
target transactional functions which in turn requires a lookup of the
transactional clone of that function. Consequently, virtual function calls are
already covered by the lookup mechanism above.

The design decisions made so far establish a sandboxing approach, which is less
intrusive then the approach of Dalessandro and Scott and covers all application
visible anomalies of our error model. The only issue still not addressed are
non-transactional sections inside transactions such as waivered sections. While
pre-validation of waivered sections would be enough for single-threaded
transactions, they will cause inconsistencies in transactions with internal
parallelism. Because we consider helper threads to execute a clone of the same
transaction, we have to deal with some additional issues in terms of
parallelised transaction in respect to waivered sections.

Waivered sections are not instrumented by definition, which implicitly results
in the following two restrictions when used with a common STM implementation:

\begin{enumerate}
  \item Waivered sections shall not access shared data accessed by transactions.
  \item Waivered sections shall not have any side effects visible to other
  threads until the transaction was committed.
\end{enumerate}

Of course, shared data can be inconsistent when read and uninstrumented writes
may violate the isolation of transactions (depending on the validation method
used by the STM implementation), which explains the first restriction. But there
are other issues to worry about: Because instrumented write access of the
transaction to shared data is buffered in the write set, uninstrumented reads to
shared data inside the waivered section will not reflect the writes of the
transaction itself, which makes their use error prone.

Regarding the second restriction, the danger of side effects visible to other
threads is also obvious as well as side effects not invisible to other threads
can be granted, too. However, some STM systems consider waivered sections to be
uninterruptible. This allows even side effects that are actually visible to
other threads as long as they are \textit{legal} in terms of intended
application behaviour. For example acquiring a lock and modifying data in the
memory protected by the lock is legal as long as the lock is released before the
transaction is left and the modification is guaranteed to be consistent. This is
easily achieved when the waivered section is uninterruptible and modifications
are reverted by some compensating action in case of a rollback. An example is
transaction-aware memory management. The allocation function uses a lock to
protect its internal data structures and the transaction keeps track of allocations to release them
in case of a rollback.

Out-of-band validation actually violates the second restriction
for waivered sections that are considered to be uninterrupted as described
above. An inconsistency detected by the out-of-band validation can
cause the interruption of the transaction and an abort inside
the waivered section. Referring to the memory allocation example above, the
lock will be left acquired and internal data of the memory management
subsystem will be inconsistent. Thus, the out-of-band validation has to be
suspended if we want to support such kind of waivered sections.

Waivered sections actually allow to work efficiently on private data. For
example a thread might have just allocated a large data structure and started a
transaction to initialise it from shared data and publish it afterwards. If the
transaction is executed by multiple threads at once, for example using a helper
thread running a cloned transaction for out-of-band validation as described
above, both threads will directly access the same data structure concurrently
inside the waivered section. While one thread may commit the other might still
run on an inconsistent state, directly writing inconsistent values to the data
structure considered private.

Although, properly applied waivered sections provide significant improvements to
the response time of transactions in general, they are error prone and complex
in use especially in terms of maintainability. This might also be the reason why
waivered sections are still not part of the proposed language extensions for
C++. Therefore, our sandboxing approach does not support waivered sections with
side effects on application level. But we do suspend the out-of-band validation for
internal purposes such as modifications to internal data of the transaction
(e.g. read set and write set) and the memory management functions supported by
the STM as well.

\section{Evaluation}

The sandboxing concept given in Section \ref{sec:CONCEPT} has been evaluated on
prototypes with different types of out-of-band validation implemented on top of
NOrec \cite{bib:NOREC} plus support for GCC 4.7 as TM aware compiler. NOrec is a
timer-based STM using deferred updates and eager validation, which was modified
by us to perform lazy validation in the prototypes and adapted to the
application binary interface \cite{bib:TMABI} of the GCC for STM library
integration. The prototypes have been compared to each other and to the original
NOrec by means of averaged total execution times of benchmarks of the STAMP
benchmark suite. The benchmarks have been executed on a 64 core system with 4
AMD Opteron\texttrademark 6282 SE processors at 3GHz clock rate, 128 GB main
memory and Debian operating system with Linux kernel version 3.2.57-3. Table
\ref{tbl:algos} lists the names used for the different STM implementations in
the evaluation.

\begin{table}
\centering
\begin{tabular}{|l|p{10.5cm}|}
\hline
\texttt{NOrec} & Original NOrec implementation without modifications.\\\hline
\texttt{NOrecSb} & NOrec with sandboxing (lazy validation) and timer-triggered
validation.\\\hline \texttt{NOrecHt} & Same as NOrecSb with a helper thread,
which validates the read set of its leader.\\\hline 
\texttt{NOrecCIV} & Same as NOrecSb with out-of-band validation in the helper
thread running a cloned transaction of its leader with eager validation.\\\hline
\end{tabular}
\caption{STM implementations}
\label{tbl:algos}
\end{table}

The benchmarks have different characteristics regarding the application of
transactions summarised in Table \ref{tab:STAMP_CHARACTERISTICS}. 
The runtime properties given in the table have the following meaning:
\begin{description}
  \item [\textbf{Tx Length:}] This property provides a rough estimation of the
  amount of instructions to be processed inside a transactional section on
  application level.
  \item [\textbf{R/W Set:}] This property gives an estimation of the average
  length of the read and write set of a transaction, which roughly reflects the
  amount of read and write operations.
  \item [\textbf{Tx Time:}] This property gives an estimation of the percentage
  of execution time spent in transactions.
  \item [\textbf{Contention:}] This property reflects the probability of
  conflicts between threads derived from the average number of aborts per
  commit.
\end{description}

\begin{table}[h!]
\begin{center}
\footnotesize
\begin{tabular}{|l|l|l|l|l|}
\hline
\textbf{Benchmark} & \textbf{Tx Length} & \textbf{R/W Set} & \textbf{Tx Time}
& \textbf{Contention} \\\hline
genome & Medium & Large & High & Low \\\hline
intruder & Short & Medium & Medium & High \\\hline
kmeans & Short & Medium & Low & Low \\\hline
labyrinth & Long & Large & High & High \\\hline
ssca2 & Short & Small & Low & Low \\\hline
vacation & Medium & Large & High & Low/Medium \\\hline
\end{tabular}
\caption{Characteristics of the STAMP benchmarks}
\label{tab:STAMP_CHARACTERISTICS}
\end{center}
\end{table}

\begin{figure}[ht]
\begin{minipage}[b]{0.4925\linewidth}
\centering
\includegraphics[width=\textwidth]{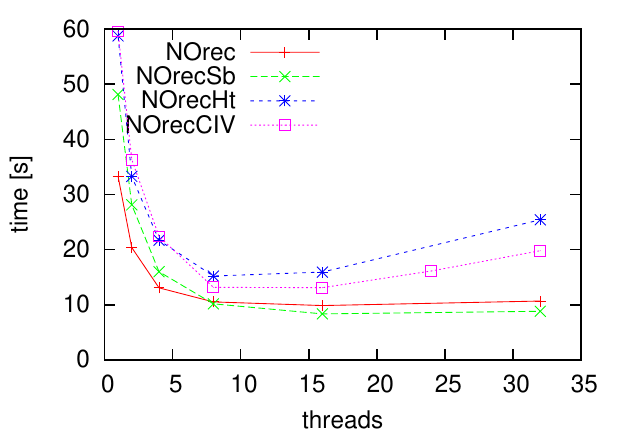}
\caption{Genome benchmark}
\label{fig:genome}
\end{minipage}
\begin{minipage}[b]{0.4925\linewidth}
\centering
\includegraphics[width=\textwidth]{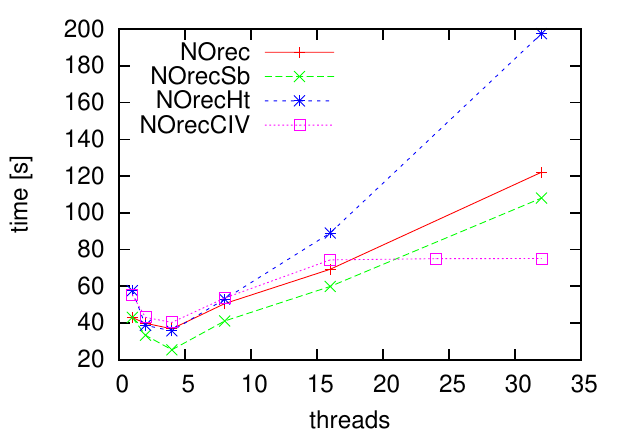}
\caption{Intruder benchmark}
\label{fig:intruder}
\end{minipage}
\end{figure}

\begin{figure}[ht]
\begin{minipage}[b]{0.49\linewidth}
\centering
\includegraphics[width=\textwidth]{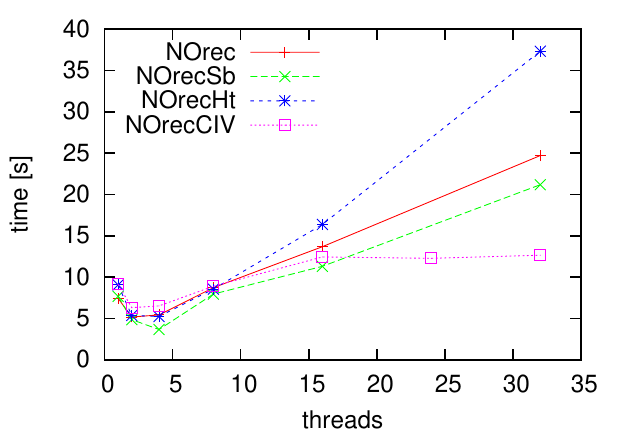}
\caption{Kmeans benchmark}
\label{fig:kmeans}
\end{minipage}
\hspace{0.1cm}
\begin{minipage}[b]{0.49\linewidth}
\centering
\includegraphics[width=\textwidth]{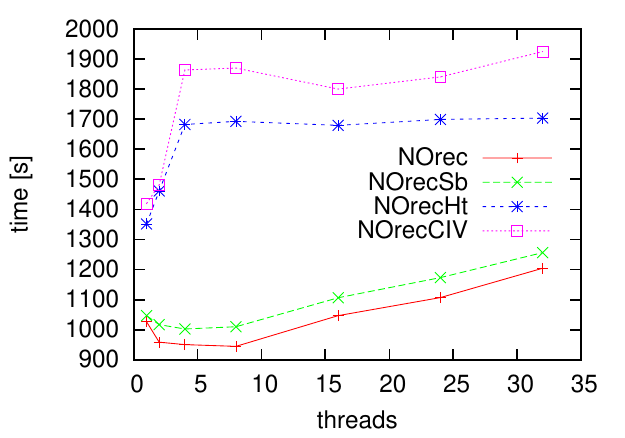}
\caption{Labyrinth benchmark}
\label{fig:labyrinth}
\end{minipage}
\end{figure}

\begin{figure}[ht]
\begin{minipage}[b]{0.49\linewidth}
\centering
\includegraphics[width=\textwidth]{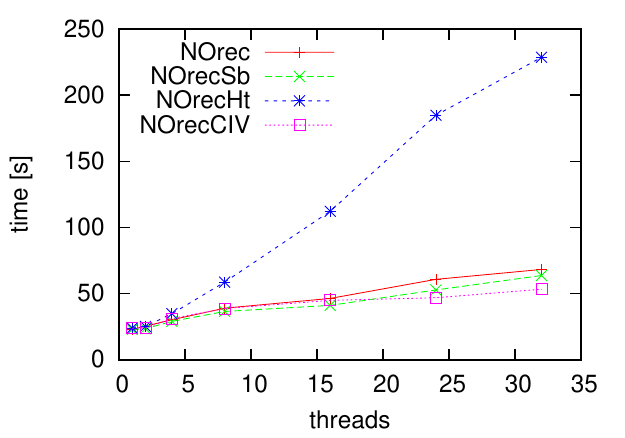}
\caption{SSCA2 benchmark}
\label{fig:ssca2}
\end{minipage}
\hspace{0.1cm}
\begin{minipage}[b]{0.49\linewidth}
\centering
\includegraphics[width=\textwidth]{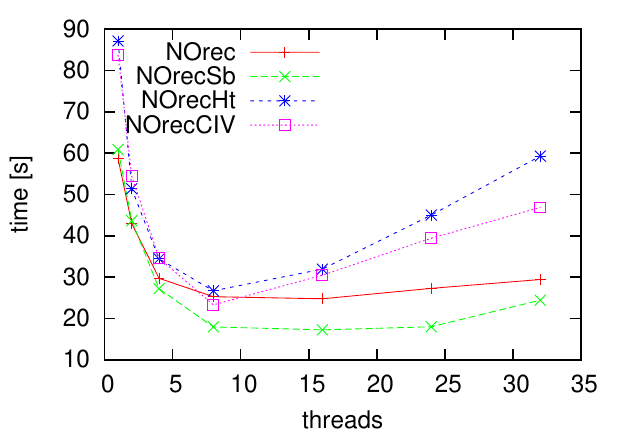}
\caption{Vacation benchmark}
\label{fig:vacation}
\end{minipage}
\end{figure}

The graphs depicted in Figures \ref{fig:genome} to \ref{fig:vacation}
contain the average response times (total execution time for a single benchmark
run) in seconds (vertical axis) with sets of $1-32$ application threads
(horizontal axis). Helper threads are not considered as application threads in
this regard. Measurement runs have been repeated multiple times for each test
configuration until the size of the 90\% confidence interval for the calculated
average fell below $5\%$ of the standard deviation.

Considering the different approaches the following results have been expected:
\texttt{NOrecHt} should perform better or equal to \texttt{NOrecSb} because the
frequency of validations is similar but the \texttt{NOrecHt} approach does its
first validation right after the initialisation of the helper thread.
\texttt{NOrecHt} should have a slight advantage in longer transactions with low
contention because the leader thread runs without the interruptions occurring
through the timer-driven validation in \texttt{NOrecSb}. The response times of
\texttt{NOrecCIV} should range between those of \texttt{NOrecSb} and the
original \texttt{NOrec} because the leader runs uninterrupted and the eager
validation of the helper should have the same reaction latency as the eager
validation in the original \texttt{NOrec} implementation.

 \begin{figure}[ht]
\centering
\includegraphics[width=\textwidth]{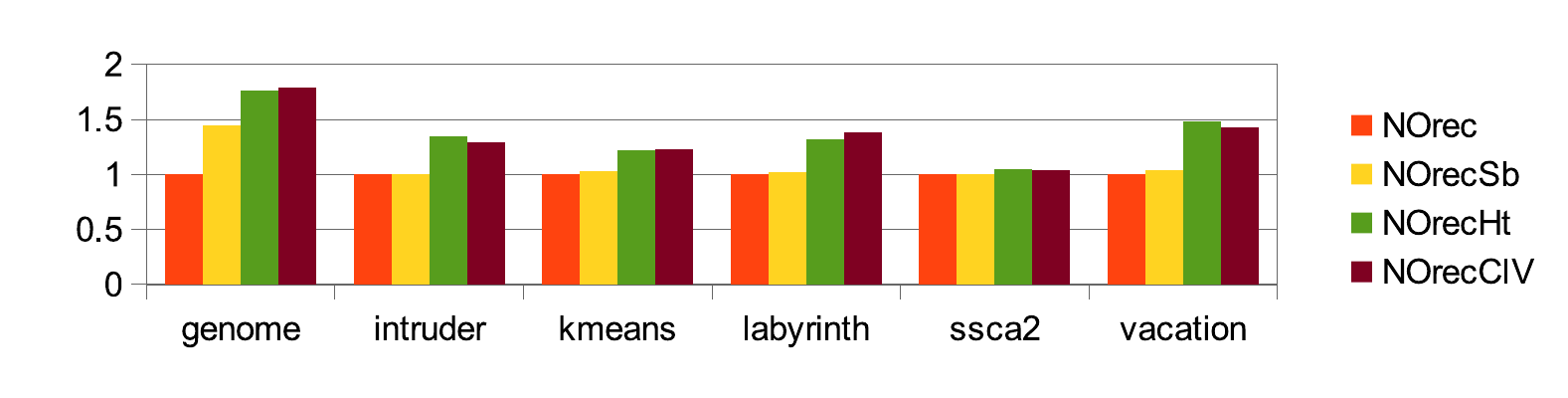}
\caption{Single-threaded overhead relative to the original NOrec}
\label{fig:single-threaded}
\end{figure}

Figure \ref{fig:single-threaded} shows the overhead in single-threaded operation
of the benchmarks in relation to the original NOrec implementation.
It clearly shows that the helper thread aided approaches (\texttt{NOrecHt} and
\texttt{NOrecCIV}) have a general disadvantage due to the effort to clone a
transaction at each transaction entry. Thus, to beat \texttt{NOrec} and
\texttt{NOrecSb} these prototypes have to compensate the cloning effort first.

The \texttt{NOrecSb} variant demonstrates the general advantage over eager
validation (original \texttt{NOrec}) in most cases. Considering just the
algorithmic behaviour and thereby ignoring the hardware influences, \texttt{NOrecHt} 
should perform similar to \texttt{NOrecSb}. But obviously, it cannot compete
with any of the other prototypes in most cases. The
observed effect gets more distinct with increased number of threads.

The frequency in which the helper thread validates the read set of the leader
is reduced to the frequency also used by the timer-driven validation in the
\texttt{NOrecSb} implementation. Thus, the amount of reads on data shared 
by all transactions is almost the same in \texttt{NOrecHt} and
\texttt{NOrecSb}. But there is additional contention between the leader and the
helper thread by sharing the read set of the leader and some synchronisation to
guarantee the consistency of the read set observed by the helper. Also, the
application data shared between the application threads is now accessed by twice
the amount of actual threads, which causes additional contention on hardware
layer. Together, these are the reasons why the \texttt{NOrecHt} variant performs
worse than \texttt{NOrecSb} on the given hardware.

\texttt{NOrecCIV} was developed especially to solve the issues with the
increased cache contention experienced with the \texttt{NOrecHt} approach. The
negative effect of the \texttt{NOrecHt} approach was dramatically reduced, which
has improved the response time significantly. Additionally, the response time of
\texttt{NOrecCIV} is in most cases close to the original \texttt{NOrec}
implementation. The best results have been scored by \texttt{NOrecCIV} with more
than 16 threads in the benchmarks with short transactions and small to medium
sized read and write sets, namely intruder (see Figure \ref{fig:intruder}),
kmeans (see Figure \ref{fig:kmeans}) and SSCA2 (see Figure \ref{fig:ssca2}).
The advantage over \texttt{NOrecSb} seems to be odd in the first place but the
timer-driven validation is configured to occur once per transaction run and
thereby optimises non-conflicting runs. Consequently, timer-driven validation is
less frequent as the eager validation of the helper thread and the
\texttt{NOrecCIV} variant is supposed to detect inconsistencies earlier in
higher concurrency scenarios.

The worst case situation for \texttt{NOrecCIV} was found with the labyrinth
benchmark (see Figure \ref{fig:labyrinth}), which runs the longest
transactions with the largest read and write sets. A possible explanation for
this effect is the additional cache space required by the helper thread. In
fact, the required amount of cache lines is doubled in case of two threads
working on the same transaction. The lack of space in the caches leads
to additional cache misses, which can be the explanation for the bad performance
in this case. This hypothesis is backed by the fact that \texttt{NOrecHt}
performs better in this benchmark. \texttt{NOrecHt} uses no own read set and
requires less space in caches, consequently. But the general disadvantage of
increased footprint in terms of cache lines is shared by both helper thread
aided prototypes as shown in the labyrinth benchmark.

\section{Conclusion}

This work has to be considered as another step to improve the techniques of
sandboxing in unmanaged languages. As Dalessandro and Scott have already shown,
sandboxing is a promising method that deserves further research.

Our approach further reduces the amount of validations required to run
transaction with deferred updates and lazy validation. Although it is a
limitation, this particular kind of concurrency control has been proven to
perform very good in comparison to all other major approaches for STM.
Unfortunately, we could not provide runtime comparison to the sandboxing
approach of Dalessandro and Scott due to several issues related to version
incompatibilities of the sub-systems they have used.

We have demonstrated, that helper thread aided out-of-band validation can
further improve sandboxing if the helper thread executes the transaction on its
own and applies eager validation to detect conflicts on behalf of the leader
thread, which runs lazy validation. This approach reduces the amount of cache
contention usually caused by a helper thread validating the read set of the
leader instead. Those results depend on the cache infrastructure provided by
the hardware, and the advantage might turn in a disadvantage on different
hardware, which should be further investigated.

We have also identified waivered sections as an issue in regards to complexity
of software development. While transactions generally reduce the complexity by
removing the risk of deadlocks/livelocks while transparently improving
scalability in comparison to traditional use of locks, waivered sections
introduce a method difficult to cope with. Additionally, waivered sections will
behave differently depending on the STM algorithm used.

Future work should aim on further consideration of the hardware characteristics,
comparing the different out-of-band validation approaches on different multi-core
architectures. Also, a comparison to the original sandboxing approach should be
arranged to gain more confidence in respect to the new stack protection. The
concept of waivered sections should be reassessed and defined more precisely,
also in regards to types of STM algorithms that will be supported and a general
method to protect against resource exhaustion can be developed.

\bibliographystyle{splncs}
\bibliography{tm-sandboxing}

\end{document}